\documentclass[letterpaper, 11pt]{article}

\newif\ifLM
\usepackage{mathtools}  % also loads `amsmath` package
\usepackage{amssymb}
\usepackage{amsthm}  % load before `newpxmath` package

\theoremstyle{plain}
\newtheorem{theorem}{Theorem}
\newtheorem{corollary}{Corollary}[theorem]

\theoremstyle{remark}
\newtheorem*{claim}{Claim}

\ifLM
\usepackage{lmodern}
\else
\usepackage[theoremfont, trueslanted, largesc]{newpxtext}  % package author wants this loaded after `babel`, but this doesn't seem necessary
\usepackage[vvarbb, noamssymbols]{newpxmath}  % package author has `fontenc` loaded before this, but it's unclear if this is necessary
\fi

\usepackage[T1]{fontenc}  % allows LaTeX to output non-ASCII characters better; load after fonts if possible
\usepackage[english]{babel}  % sets language to American English
\AddBabelHook{frspacing}{afterextras}{\frenchspacing}  % enables single sentence spacing

\usepackage[hmargin=1.2in, marginparwidth=1in, marginparsep=0.1in, vmargin=1in, headheight=0.6in, headsep=0.3in, footskip=0.5in, heightrounded]{geometry}  % useful for setting page dimensions and layout

\usepackage{ftcap}  % fixes table–caption spacing in `article` class
\usepackage[numbers, sort&compress]{natbib}  % improves formatting of citations and references
\usepackage[tracking]{microtype}  % improves typographic details (e.g., kerning)
\usepackage{fnpct}  % adjusts kerning of footnote marks (superscripts) before commas or periods

\usepackage{graphicx}  % allows inclusion of images in document
\graphicspath{{./}}  % location of images in file system

\usepackage{booktabs}
\allowdisplaybreaks[1]  % allows page breaks within `amsmath` environments; optional value sets permissibility (min=1, max=4)

\ifLM\fi  % use for matrix transpose

\usepackage{bm}  % use `\bm` for bold math symbols

\newcommand*{\phanrel}{\mathrel{\phantom{=}}}
\newcommand*{\Reals}{\mathbb{R}}

\DeclareMathOperator{\E}{\mathbb{E}}

\DeclarePairedDelimiterX\set[1]\lbrace\rbrace{\mkern1.5mu#1\mkern1.5mu}
\usepackage{tikz}
\usetikzlibrary{babel, positioning}

\usepackage[colorlinks, allcolors=blue]{hyperref}  % enables hyperlinks; load before `ellipsis` package
\hypersetup{
    pdftitle=Online Maximum Independent Set of Hyperrectangles,
    pdfauthor=Rishi Advani and Abolfazl Asudeh,
    pdfkeywords=Maximum independent set; hyperrectangles; geometric intersection graphs; online algorithms; competitive analysis,
    pdflang=en-US
}

\usepackage{ellipsis}  % adjusts spacing of ellipses before punctuation

\newcommand*{\detgreedy}{{\small $\textsc{DetGreedy}$}}
\newcommand*{\greedy}[1]{{\small $\textsc{Greedy}_{#1}$}}
\newcommand*{\selgreedy}[1]{{\small $\textsc{SelectiveGreedy}_{#1}$}}

\begin{document}

\title{Online Maximum Independent Set of Hyperrectangles}

\author{Rishi Advani\\University of Illinois Chicago\\radvani2@uic.edu \and Abolfazl Asudeh\\University of Illinois Chicago\\asudeh@uic.edu}

\date{}

\maketitle

\begin{abstract}
The maximum independent set problem is a classical NP-hard problem in theoretical computer science. In this work, we study a special case where the family of graphs considered is restricted to intersection graphs of sets of axis-aligned hyperrectangles and the input is provided in an online fashion. We prove results for several adversary models, classes of hyperrectangles, and restrictions on the order of the input. Under the adaptive offline and adaptive online adversary models, we find the optimal online algorithm for unit hypercubes, $\sigma$-bounded hypercubes, unit-volume hyperrectangles, and arbitrary hypercubes, in both non-dominated and arbitrary order. Under the oblivious adversary model, we prove bounds on the competitive ratio of an optimal online algorithm for the same classes of hyperrectangles and input orders, and we find algorithms that are optimal up to constant factors. For input in dominating order, we find the optimal online algorithm for arbitrary hyperrectangles under all adversary models. We conclude by discussing several promising directions for future work.
\end{abstract}

\section{Introduction}
The maximum independent set (MIS) problem is a classical NP-hard problem in theoretical computer science, with many real-world applications as well.
Many variations of the MIS problem have been studied. Recently, there has been a series of papers studying the MIS problem restricted to the family of intersection graphs of sets of axis-aligned rectangles in the plane~\citep{Mitchell2021, Galvez2021}.
In this work, we study the extension of this problem to higher dimensions in an online setting.

\subsection{Definitions}
In $d$ dimensions, an \emph{axis-aligned hyperrectangle} is a closed region $[a_1, b_1] \times [a_2, b_2] \times \dots \times [a_d, b_d]$, where $a_i$, $b_i \in \Reals$ for all $i$. A $d$-hyperrectangle is a $d$-dimensional hyperrectangle (e.g., a 2-hyperrectangle is a rectangle).
All hyperrectangles henceforth are assumed to be axis aligned unless otherwise specified.
A \emph{hypercube} is a hyperrectangle with all side lengths equal (i.e., $b_1 - a_1 = b_2 - a_2 = \dots = b_d - a_d$). A $d$-hypercube is a $d$-dimensional hypercube (e.g., a 3-hypercube is a cube).
Regardless of the dimension $d$, we will use ``volume'' to refer to the Lebesgue measure of a given hyperrectangle (e.g., for $d=1$, this would be the length; for $d=2$, the area).

Every hyperrectangle has exactly one vertex with all coordinates greater than or equal to the respective coordinates of every other vertex; we will refer to this vertex as the \emph{upper} vertex. Analogously, we will refer to the vertex with all coordinates less than or equal to the respective coordinates of every other vertex as the \emph{lower} vertex of the hyperrectangle. E.g., for the hypercube $[0,1]^4$, the upper vertex is $(1,1,1,1)$ and the lower vertex is $(0,0,0,0)$.
A hyperrectangle $x$ \emph{dominates} a hyperrectangle $y$ if the upper vertex of $x$ has all coordinates greater than or equal to the respective coordinates of the upper vertex of $y$. E.g., $[0,2]^2$ dominates $[0,1]^2$, but neither of $[0,1] \times [0,2]$ and $[0,2] \times [0,1]$ dominates the other.

Given a set of geometric objects (e.g., hyperrectangles) in space, their intersection graph is a simple, undirected graph with a node representing each object and an edge between the nodes of two objects that intersect.

For any (possibly randomized) online algorithm with solution $\mathrm{SOL}$, if the solution of an optimal offline algorithm is $\mathrm{OPT}$ and the space of possible inputs is denoted by $\mathcal{X}$, the \emph{competitive ratio} of the online algorithm is $\sup_{X \in \mathcal{X}} \frac{|\mathrm{OPT}|}{\E[|\mathrm{SOL}|]}$.

\subsection{Problem Formulation}
Given a set of $n$ axis-aligned hyperrectangles provided in an online fashion, the goal is to select a set of disjoint hyperrectangles with (approximately) maximum cardinality\footnote{An overview of some of the applications of this problem can be found in Appendix~\ref{sec:applications}.}.
The input is insertion only: exactly one hyperrectangle is given as input in each time step, and no hyperrectangle is ever removed. Furthermore, the selections of the algorithm are final. In each time step, the algorithm either accepts or rejects the currently offered hyperrectangle; its decision cannot be revoked in future time steps.

We prove results for various orderings of hyperrectangles in the input. The following input orders are considered (in increasing order of difficulty):
\begin{description}
    \item[Dominating] Each hyperrectangle dominates every hyperrectangle preceding it.
    \item[Non-dominated] Each hyperrectangle is \emph{not} dominated by any hyperrectangle preceding it.
    This ordering has ties to Pareto efficiency and the skyline operator~\citep{Borzsony2001}.
    \item[Arbitrary] The input can arrive in any order.
\end{description}
We also prove results under different adversary models. The following models are considered (in increasing order of difficulty):
\begin{description}
	\item[Oblivious] This adversary knows the algorithm's implementation but not the hyperrectangles it selects during execution.
	\item[Adaptive online] This adversary knows the algorithm's implementation and can adapt its strategy after the algorithm makes selections.
	\item[Adaptive offline] This adversary knows \emph{everything}. It knows the algorithm's implementation and the outcome of every random variable used in its selection process in advance.
\end{description}

Finally, we define a specific online algorithm that will be referenced throughout the paper. \detgreedy{} is defined as the algorithm that accepts every hyperrectangle given as input as long as it doesn't overlap with previously accepted hyperrectangles.

\subsection{Contributions}
To the best of our knowledge, we are the first to present results on the online MIS of hyperrectangles problem under the oblivious adversary model. We are also the first to present results under adaptive adversary models with input in dominating or non-dominated order. For input in arbitrary order, we present the first results on hyperrectangles more general than unit hypercubes; we present the first results on $\sigma$-bounded hypercubes, unit-volume hyperrectangles, and arbitrary hypercubes.

An overview of our key results for input in non-dominated and arbitrary order can be found in Tables~\ref{tab:non-dom_results}~and~\ref{tab:arbit_results}, respectively\footnote{We consider $d$ to be a constant, so all of our competitive ratios are at most polynomial.}. Under both adaptive adversary models, we show that \detgreedy{} is the optimal online algorithm. For input in dominating order, we show that the performance of \detgreedy{} matches the performance of an optimal offline algorithm for arbitrary hyperrectangles; i.e., \detgreedy{} is the best possible algorithm.

Under the oblivious adversary model, we give lower bounds on the competitive ratios of a hypothetical optimal online algorithm and the \greedy{p} algorithm (defined in \S~\ref{sec:unit:non-dominated:oblivious}). For $\sigma$-bounded hypercubes, we propose the \selgreedy{k} algorithm (defined in \S~\ref{sec:sig-bounded:non-dominated:oblivious}) and show that it performs better than \detgreedy{} (competitive ratio of $\Theta(\log(\sigma))$ vs.\ $\Omega(\sigma^{d-1})$). We prove that, up to constant factors, \selgreedy{k} is optimal for $\sigma$-bounded hypercubes and \detgreedy{} is optimal for unit hypercubes, unit-volume hyperrectangles, and arbitrary hypercubes.

We also present experimental analysis of the problem; this can be found in Appendix~\ref{sec:experiments}.

\begin{table}[bhpt]
\caption{Competitive ratios (bounds when unknown) of optimal online algorithm for input in non-dominated order.}
\label{tab:non-dom_results}
\centering
\begin{tabular}{lcc}
    \toprule
    Input shape/size & Adaptive offline/online & Oblivious \\
    \midrule
    Unit hypercubes & $2^d - 1$ & $\Bigl[ \frac{12}{7}, 2^d - 1 \Bigr]$ \\
    $\sigma$-bounded hypercubes & $(\lceil \sigma \rceil + 1)^d - \lceil \sigma \rceil^d$ & $\Bigl[ \frac{\lceil \log_2 \sigma \rceil + 1}{2}, 3^d \lceil \log_2 \sigma \rceil - 2^d \lceil \log_2 \sigma \rceil \Bigr]$ \\
    \bottomrule
\end{tabular}
\end{table}

\begin{table}[bhpt]
\caption{Competitive ratios (bounds when unknown) of optimal online algorithm for input in arbitrary order.}
\label{tab:arbit_results}
\centering
\begin{tabular}{lcc}
    \toprule
    Input shape/size & Adaptive offline/online & Oblivious \\
    \midrule
    Unit hypercubes & $2^d$~(\citet{De2021}) & $\Bigl[\frac{32}{15}, 2^d\Bigr]$ \\
    $\sigma$-bounded hypercubes & $(\lceil \sigma \rceil + 1)^d$ & $\Bigl[ \frac{\lceil \log_2 \sigma \rceil + 1}{2}, 3^d \lceil \log_2 \sigma \rceil \Bigr]$ \\
    \bottomrule
\end{tabular}
\end{table}

% \begin{table}[bhpt]
% \caption{Lower bounds on competitive ratios of \greedy{p} for $d \geq 2$ in the oblivious adversary model. The parameter $p$ gives the probability of selecting each non-overlapping hyperrectangle.}
% \label{tab:greedy_p_results}
% \centering
% \begin{tabular}{lcc}
%     \toprule
%     Input shape/size & Non-dominated order & Arbitrary order \\
%     \midrule
%     Unit hypercubes & 2.65 ($p = 0.56$) & 3.20 ($p = 0.50$) \\
%     Unit-volume hyperrectangles & 3.20 ($p = 0.50$) & 3.20 ($p = 0.50$) \\
%     \bottomrule
% \end{tabular}
% \end{table}

\subsection{Organization}
We begin by surveying related work (\S~\ref{sec:related_work}). Then, we give an analysis of the dominating order case (\S~\ref{sec:dominating}). Next, we present our results on unit hypercubes (\S~\ref{sec:unit}), $\sigma$-bounded hypercubes (\S~\ref{sec:sig-bounded}), unit-volume hyperrectangles (\S~\ref{sec:unit-vol}), and arbitrary hypercubes (\S~\ref{sec:cube}). Finally, we discuss directions for future work (\S~\ref{sec:future_work}).

\section{Related Work}
\label{sec:related_work}
In this section, we discuss key related work. Broadly, each paper can be classified as solving online MIS for different class of graphs or solving MIS of hyperrectangles under a different input framework (offline, dynamic, etc.).

\subsection{Online Interval Scheduling}
The 1-dimensional version of our problem is ``online MIS in interval graphs'' or ``online maximum disjoint set of intervals''. It can also be viewed as a form of interval scheduling.
A prominent early work on a similar problem is that of \citet{Lipton1994}. However, they aim to find a set of intervals of maximum total length instead of a set of maximum cardinality, and intervals are added in order of start time. A special case of the work of \citet{Faigle1995} is similar to our problem; they do optimize for maximum cardinality, but still require that intervals arrive in order of start time and also allow intervals to be discarded.

Most papers tend to assume that the intervals are added in order of start or end time, but \citet{Coffman1998} study the case of unit-length intervals added in random order. The model of \citet{Goyal2020} can be viewed as intervals added in arbitrary order (but maximizing weight, not cardinality).
There is also work on the streaming setting (limited memory)~\citep{Emek2016, Cabello2017}.

\subsection{Online MIS in Disk Graphs}
A disk graph is the intersection graph of a set of disks in the plane.
There is a deterministic (greedy) online algorithm for MIS in disk graphs. In unit disk graphs (all disks have diameter 1), the algorithm is 5-competitive. In general disk graphs, the algorithm is $(n-1)$-competitive. In $\sigma$-bounded disk graphs (all disks have diameters between 1 and $\sigma$), the algorithm is $O(\min\set{n, \sigma^2})$-competitive. In all cases, it is optimal~\citep{Erlebach2006}.

\citet{Caragiannis2007} study randomized algorithms for online MIS in disk graphs. They use the oblivious adversary model. They give a lower bound of $\Omega(n)$ on the competitive ratio for general disks. For $\sigma$-bounded disks, they give an algorithm that is $O(\min\set{n, \log\sigma})$-competitive if $\sigma$ is given as input and an algorithm that is $O\bigl(\min\set[\big]{n, \prod_{j=1}^{\log^* \sigma-1} \log^{(j)} \sigma}\bigr)$-competitive otherwise. For unit disk graphs, they give a lower bound of $2.5$ (or 3 if only the graph representation is given) and present an algorithm with a competitive ratio of $8 \sqrt{3} / \pi \approx 4.41$.
For the similar case of unit balls, there is a 12-competitive algorithm~\citep{De2021}.
In the (turnstile) streaming case for unit disks, \citet{Bakshi2020} prove the same $8 \sqrt{3} / \pi$ upper bound and also prove a $2-\epsilon$ lower bound.

\citet{Marathe1995} show that there is a 3-competitive algorithm in unit disk graphs where disks arrive in order of $x$-position. They also mention the idea of inscribing regular polygons in unit disks to get algorithms with loose competitive ratios for MIS for that type of polygon. One could use this approach to compute loose competitive ratios for unit squares arriving in order of $x$-position. Furthermore, the same idea could potentially be extended to regular polygons (squares) inscribed in general disks, irregular polygons (rectangles) inscribed in unit disks, or irregular polygons (rectangles) inscribed in general disks.

\subsection{Online MIS in General Graphs}
For certain stochastic input models, there is an algorithm with competitive ratio quadratic in the inductive independence number\footnote{A graph has inductive independence number at most $\rho$ if the vertices can be ordered such that for every independent set $S$ and vertex $v$, at most $\rho$ vertices in $S$ simultaneously neighbor and are preceded by $v$.} of the graph~\citep{Gobel2013}.
If the algorithm is allowed to maintain multiple candidate solutions, then there is an algorithm with sublinear competitive ratio~\citep{Halldorsson2002}.
Other relaxations of the online setting have been considered as well~\citep{Boyar2022}.

\subsection{Offline MIS of Rectangles}
The offline 2-dimensional version of our problem has received a lot of recent attention. To the best of our knowledge, \citet{Agarwal1998} were the first to study the problem. The first (polynomial-time) constant-factor approximation algorithm was only recently discovered by \citet{Mitchell2021}. The current state of the art has a competitive ratio of $2 + \epsilon$~\citep{Galvez2021}.
There is also existing work where restrictions are placed on the family of rectangles used as input~\citep{Correa2015}.

\subsection{Dynamic/Streaming/Online MIS of Hyperrectangles}
The dynamic version of MIS of hyperrectangles has been studied before~\citep{Bhore2022a}. There are also earlier results for hypercubes in a dynamic setting~\citep{Bhore2020}.
\citet{Bhore2022b} study MIS of rectangles in a streaming setting (limited memory). They present a one-pass streaming algorithm that achieves an approximation factor of 4 for unit-height rectangles.

There is a single work studying a subcase of the problem we study in this paper. \citet{De2021} have independently proved tight upper and lower bounds for the special case of unit $d$-cubes in arbitrary order with an adaptive adversary.
In addition, the inductive independence number of intersection graphs of $d$-hyperrectangles is $2d-1$~\citep{Ye2012}, so there is an algorithm with constant competitive ratio for online MIS of $d$-hyperrectangles for certain stochastic input models~\citep{Gobel2013}.

Finally, since this paper's initial submission, a paper has been made available on arXiv studying the online MIS of hyperrectangles problem, but with the input arriving in random order~\citep{Garg2024}.

\section{Dominating Order}
\label{sec:dominating}
The order in which the input arrives has a large impact on the hardness of the problem. In this section, we show that \detgreedy{} finds an optimal solution even with arbitrary hyperrectangles under the adaptive offline adversary model when the input arrives in dominating order.

\begin{theorem}
    \detgreedy{} is optimal.
\end{theorem}

\begin{proof}
	Note that a hyperrectangle can be intersected by any number of disjoint hyperrectangles if the input comes in dominating order, but it can be dominated by at most one of those and dominates the rest.
	Let $A$ be the set of hyperrectangles in some optimal solution. Let $B$ be the set of hyperrectangles in the greedy solution.
	Let $x_i \in B \setminus A$. Consider the set $Y_i$ of hyperrectangles in $A \setminus B$ that intersect $x_i$ and are dominated by it. Since each hyperrectangle in $Y_i$ was not selected by \detgreedy{}, they must each intersect some hyperrectangle in $B$ other than $x_i$.
	
	If $x_i$ is the first hyperrectangle in the input order, then $|Y_i| = 0$. Otherwise, let $z_i$ be the hyperrectangle in $B$ whose position in the input order is closest to that of $x_i$ out of those that came before $x_i$. Note that every hyperrectangle in $Y_i$ must intersect $z_i$. Since the hyperrectangles in $Y_i \subset A$ must be disjoint, $Y_i$ must have cardinality at most 1.
	
	Thus, each hyperrectangle $x_i$ in $B \setminus A$ intersects at most two hyperrectangles in $A \setminus B$. If it intersects two, then one of them dominates it and the other is dominated by it. Furthermore, the hyperrectangle that is dominated by $x_i$ must intersect (and dominate) some other hyperrectangle in $B \setminus A$ as well. Therefore, $|A| = |B|$.
\end{proof}

Essentially, in dominating order, it reduces to the 1-dimensional case, where ``earliest deadline first'' is optimal.

\section{Unit Hypercubes}
\label{sec:unit}
We now begin our analysis of hyperrectangles in non-dominated and arbitrary order. In this section, we focus on unit hypercubes, the most restrictive class of hyperrectangles considered.

\subsection{Non-Dominated Order and Adaptive Adversaries}
\begin{theorem}
    \detgreedy{} has a competitive ratio of at most $2^d-1$ under the adaptive offline adversary model.
\end{theorem}

\begin{proof}
	Note that a unit $d$-cube can be intersected by at most $2^d$ disjoint unit $d$-cubes if the input comes in non-dominated order. Let $A$ be the set of hypercubes in some optimal solution. Let $B$ be the set of hypercubes in the greedy solution. For each hypercube $x_i$ in $B \setminus A$, there are at most $2^d$ hypercubes in $A \setminus B$ that intersect it. If there is one, let $y_i$ be the hypercube that is dominated by $x_i$. Since $y_i$ was not chosen by \detgreedy{}, it must intersect some other hypercube in $B$. Consequently, $A \setminus B$ can be partitioned into subsets of size at most $2^d - 1$, each subset associated with a distinct hypercube in $B \setminus A$. Thus, we have the following.
	\begin{align*}
		|A \setminus B| &\leq (2^d - 1) \cdot |B \setminus A| \\
		|A \setminus B| + |A \cap B| &\leq (2^d - 1) \cdot |B \setminus A| + |A \cap B| \\
		|A \setminus B| + |A \cap B| &\leq (2^d - 1) \cdot |B \setminus A| + (2^d - 1) \cdot |A \cap B| \\
		|A \setminus B| + |A \cap B| &\leq (2^d - 1) \cdot ( |B \setminus A| + |A \cap B| ) \\
		|A| &\leq (2^d - 1) \cdot |B|
	\end{align*}
\end{proof}

\begin{theorem}
    Every online algorithm must have competitive ratio at least $2^d - 1$ under the adaptive online adversary model.
\end{theorem}

\begin{proof}
	Until the algorithm selects a hypercube, the adversary gives as input disjoint, unit hypercubes. Once the algorithm selects a hypercube $x_i$, the adversary gives as input (in non-dominated order) $2^d - 1$ disjoint, unit hypercubes that intersect $x_i$ but are not dominated by $x_i$.
\end{proof}

\begin{corollary}
    \detgreedy{} is optimal for unit hypercubes in non-dominated order under adaptive adversary models.
\end{corollary}

\subsection{Non-Dominated Order and an Oblivious Adversary}
\label{sec:unit:non-dominated:oblivious}
Our proposed algorithm is ``greedy with probability $p$'': if the newest hyperrectangle in the input does not overlap with previously accepted hyperrectangles, accept it with probability $p$.
For any $p \in [0,1]$, we will denote this algorithm by \greedy{p}.
If the adversary makes none of the input overlap, this gives us a ratio of $1/p$, so the competitive ratio is at least $1/p$ for any $p$.
If the adversary repeatedly gives a $d$-cube and then $2^d - 1$ disjoint hypercubes that intersect it (as in the adaptive case), then the ratio is
\[\frac{2^d - 1}{p + (1-p) (2^d - 1) p} = \frac{2^d - 1}{2^d p - (2^d - 1) p^2} \,.\]
Differentiating the denominator, we have $2^d - (2^{d+1} - 2) p$, which has a zero at $p = \frac{2^{d-1}}{2^d - 1}$, giving us a ratio of $\frac{(2^d - 1)^2}{4^{d-1}}$. Thus, the competitive ratio is at least $\frac{(2^d - 1)^2}{4^{d-1}}$ for $n = 2^d$ and $p = \frac{2^{d-1}}{2^d - 1}$. This ratio is approximately 4 for large $d$.

For $n$ hypercubes, there are $2^{T_{n-1}}$ possible intersection graphs (accounts for the ordering as well), where $T_n$ is the $n$th triangular number, so it is likely infeasible to find the optimal strategy by hand.
We implemented a brute force approach~\citep{Advani2024} that tries all possible intersection graphs to find the optimal strategy for an oblivious adversary for each $p \in \set{0.00, 0.01, \dots, 1.00}$.
Competitive ratios for various values of $n$ and $d \geq 2$ are shown in Table~\ref{tab:upper_bounds}. For $d=1$ they are only upper bounds. They were calculated by running the code and then manually determining valid arrangements of unit squares in the plane with the corresponding intersection graphs. Arrangements for $n = 3 \dots 5$ are shown in Appendix~\ref{sec:additional_arrs}.
In addition, with $p=0.5$, for $n=6$ and $n=7$, we have upper bounds of 3.200000 and 3.636364, respectively, on the competitive ratio.
% corresponding `G_str` values:
% n=6: '111110110011000'
% n=7: '111110110011000110000'

\begin{table}[bhpt]
\caption{Competitive ratios for approximately optimal values of $p$ and $d \geq 2$.}
\label{tab:upper_bounds}
\centering
\begin{tabular}{cccc}
    \toprule
    $n$ & $p$ & competitive ratio \\% & runtime (s) \\
    \midrule
    1 & 1.00 & 1.000000 \\% & $<1$ \\
    2 & 1.00 & 1.000000 \\% & $<1$ \\
    3 & 0.75 & 1.777778 \\% & $<1$ \\
    4 & 0.67 & 2.250056 \\% & 1 \\
    5 & 0.56 & 2.651001 \\% & 38 \\
    \bottomrule
\end{tabular}
\end{table}

If we want to marginally improve these results, we can analytically find the optimal $p$ for each intersection graph and then confirm by brute force that the same graph is optimal for that new $p$. For example, for $n=5$, we need to find the $p$ that maximizes the following expression, where $X$ is the number of hypercubes selected.
\begin{align*}
	\E[X]
    &\begin{multlined}[t]
	= p + (1-p) (p + (1-p) (p \cdot (1 + p \cdot (1 + p) + (1-p) p) \\
    + (1-p) (p \cdot (1 + p) + (1-p) p)))
    \end{multlined} \\
	&= p + (1-p) (p + (1-p) (p \cdot (1 + 2p) + (1-p) 2p)) \\
	&= p + (1-p) (p + (1-p) 3p) \\
	&= p + (1-p) (4p - 3p^2) \\
	&= p + 4p - 3p^2 - 4p^2 + 3p^3 \\
	&= 5p - 7p^2 + 3p^3
\end{align*}
Differentiating, we have $\frac{d}{dp} \E[X] = 5 - 14p + 9p^2$, which has zeros at $\frac{5}{9}$ and $1$. The expectation is maximized at $p = \frac{5}{9}$. The code confirms that the optimal intersection graph for this value of $p$ is the same, so the competitive ratio of \greedy{\frac{5}{9}} for $n=5$ is $\frac{729}{275} \approx 2.650909$.

For general $n>5$ and $d \geq 2$, the adversary's strategy must be at least as good as using the optimal strategy for $n=5$ for each set of 5 hypercubes and then using the optimal strategy for $n \bmod 5$ for the remaining hypercubes. If we fix $p=0.56$, we have an asymptotic lower bound of 2.651001 for the competitive ratio.

One difficulty with this computational approach is that we only get the optimal intersection graph, not the arrangement of the hypercubes in space. This means that we have to manually verify if there even exists such an arrangement of unit $d$-cubes corresponding to the graph. On the other hand, even though this approach does not guarantee the existence of the actual arrangement of unit hypercubes, it does give us an upper bound on the competitive ratio.

Another obstacle is that the expression for the expected value of the solution constructed by \greedy{p} is given by a polynomial of degree $n$. For large $n$, this could lead to huge changes in the value of the expression with very small changes to $p$. This won't invalidate our theoretical results, but it does mean we may miss even better algorithms given by other values of $p$ if we are not very careful in the optimization process. Perhaps there are some theoretical guarantees on how many values of $p$ we have to test before we can be reasonably sure that we have found the maximum? Is there a more practical method for maximizing univariate polynomials over a finite interval? These are two relevant questions for future work.

\begin{theorem}
    Every online algorithm must have competitive ratio at least $\frac{12}{7}$ asymptotically.
\end{theorem}

\begin{proof}
	The adversary begins in State $i$. The adversary first gives two disjoint unit hypercubes as input. Then, it (uniformly) randomly ``marks'' one (denote it by $x_{i,1}$) and gives as input two disjoint, unit hypercubes that intersect $x_{i,1}$ but no other hypercubes from the existing input. Next, it randomly marks one of these (denote it by $x_{i,2}$) and gives as input two disjoint, unit hypercubes that intersect $x_{i,1}$ and $x_{i,2}$ but no others. Finally, the adversary switches to State $i+1$ and restarts the process from $j=1$. Once $n$ hypercubes have been given as input (by the end of State $I \coloneqq \lceil n/4 \rceil$), there is an MIS of size $\bigl\lfloor \frac{n}{2} \bigr\rfloor + I$ consisting of all the hypercubes not marked by the adversary.
	
	Consider any online algorithm. For each state $i$ and level $j$, by symmetry, the probability that it selects only $x_{i,j}$ and the probability that it selects only the other hypercube are equal; denote this probability by $p_{i,j}$. Let $p_{\sup} = \sup \set{p_{i,j}}$. Then, in each state $i \ne I$, the expected size of the independent set $X_i$ constructed by the online algorithm can be bounded as follows.
	\[\E[X_i] \leq 1 + p_{i,1} (1 + p_{i,2}) \leq 1 + p_{\sup} (1 + p_{\sup}) \leq 1 + \frac{1}{2} \biggl(1 + \frac{1}{2}\biggr) = \frac{7}{4}\]
	The expected size of the overall independent set $\bigcup \set{X_i}_{i=1}^{I}$ is then at most $\frac{7}{4} I = \frac{7}{4} \bigl\lceil \frac{n}{4} \bigr\rceil$.
	Finally, the competitive ratio is at least
	\[
	\frac{\bigl\lfloor \frac{n}{2} \bigr\rfloor + \lceil \frac{n}{4} \rceil}{\frac{7}{4} \bigl\lceil \frac{n}{4} \bigr\rceil}
	= \frac{\bigl\lfloor \frac{n}{2} \bigr\rfloor}{\frac{7}{4} \bigl\lceil \frac{n}{4} \bigr\rceil} + \frac{4}{7}
	\,,\]
	and is asymptotically at least $\frac{12}{7}$.
\end{proof}

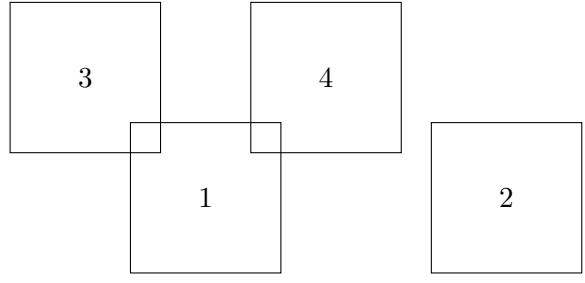
\begin{figure}[bhpt]
	\centering
	\begin{tikzpicture}[scale=1]
		\draw (2.5,2.5) rectangle node {1} (3.5,3.5);
		\draw (4.5,2.5) rectangle node {2} (5.5,3.5);
		\draw (1.7,3.3) rectangle node {3} (2.7,4.3);
		\draw (3.3,3.3) rectangle node {4} (4.3,4.3);
	\end{tikzpicture}
	\caption{An example of an oblivious adversary strategy for unit hypercubes in non-dominated order.}
	\label{fig:adversary_strategy_unit_1}
\end{figure}

An example of the adversary strategy from the proof is shown in Figure~\ref{fig:adversary_strategy_unit_1}.

\subsection{Arbitrary Order and Adaptive Adversaries}
The results in this section were proved independently by \citet{De2021}.

\begin{theorem}
    \detgreedy{} has a competitive ratio of at most $2^d$ under the adaptive offline adversary model.
\end{theorem}

\begin{proof}
	Note that a unit $d$-cube can be intersected by at most $2^d$ disjoint unit $d$-cubes. Let $A$ be the set of hypercubes in some optimal solution. Let $B$ be the set of hypercubes in the greedy solution. For each hypercube in $B \setminus A$, there are at most $2^d$ hypercubes in $A \setminus B$ that intersect it. By construction, every hypercube in $A \setminus B$ must intersect some hypercube in $B \setminus A$ (or it would be in $B$). Consequently, $A \setminus B$ can be partitioned into subsets of size at most $2^d$, each subset associated with a distinct hypercube in $B \setminus A$. Thus, we have the following.
	\begin{align*}
		|A \setminus B| &\leq 2^d \cdot |B \setminus A| \\
		|A \setminus B| + |A \cap B| &\leq 2^d \cdot |B \setminus A| + |A \cap B| \\
		|A \setminus B| + |A \cap B| &\leq 2^d \cdot |B \setminus A| + 2^d \cdot |A \cap B| \\
		|A \setminus B| + |A \cap B| &\leq 2^d \cdot ( |B \setminus A| + |A \cap B| ) \\
		|A| &\leq 2^d \cdot |B|
	\end{align*}
\end{proof}

\begin{theorem}\label{thm:unit:arbitrary:adaptive-online}
    Every online algorithm must have competitive ratio at least $2^d$ under the adaptive online adversary model.
\end{theorem}

\begin{proof}
    Until the algorithm selects a hypercube, the adversary gives as input disjoint, unit hypercubes. Once the algorithm selects a hypercube $x_i$, the adversary gives as input $2^d$ disjoint, unit hypercubes that intersect $x_i$.
\end{proof}

\begin{corollary}
    \detgreedy{} is optimal for unit hypercubes in arbitrary order under adaptive adversary models.
\end{corollary}

\subsection{Arbitrary Order and an Oblivious Adversary}
\label{sec:unit:arbitrary:oblivious}
Again, our proposed algorithm is \greedy{p}.
If the adversary makes none of the input overlap, this gives us a ratio of $1/p$. so the competitive ratio is at least $1/p$ for any $p$.
If the adversary repeatedly gives a $d$-cube and then $2^d$ disjoint hypercubes that intersect it (as in the adaptive case), then the ratio is
\[\frac{2^d}{p + (1-p) 2^d p} = \frac{2^d}{(2^d + 1)p - 2^d p^2} \,.\]
Differentiating the denominator, we have $2^d + 1 - 2^{d+1} p$, which has a zero at $p = \frac{2^d + 1}{2^{d+1}}$, giving us a ratio of $\frac{4^{d+1}}{(2^d + 1)^2}$. Thus, the competitive ratio is at least $\frac{4^{d+1}}{(2^d + 1)^2}$ for $n = 2^d + 1$ and $p = \frac{2^d + 1}{2^{d+1}}$. This ratio is approximately 4 for large $d$.

The same competitive ratios from Table~\ref{tab:upper_bounds} apply here as well. Again, for $d=1$ they are only upper bounds. In addition, for $n=6$, we have found an explicit arrangement (see Appendix~\ref{sec:additional_arrs}), so for $d \geq 2$, we can state an approximately optimal value of $p$ and competitive ratio of $0.50$ and $3.200000$, respectively.  % runtime of 2577 seconds
For $n=7$, we do not have an explicit arrangement, so the value from the non-dominated case remains just an upper bound.

For general $n>6$ and $d \geq 2$, the adversary's strategy must be at least as good as using the optimal strategy for $n=6$ for each set of 6 hypercubes and then using the optimal strategy for $n \bmod 6$ for the remaining hypercubes. If we fix $p=0.5$, we have an asymptotic lower bound of 3.2 for the competitive ratio.

\citet{Caragiannis2007} prove a result for the similar case of unit disks in arbitrary order with an oblivious adversary that beats the optimal solution for an adaptive adversary. Unfortunately, when adapted to the setting of unit hypercubes, their technique does not improve on the optimal solution for an adaptive adversary.

%proof sketch:
%
%select a random point uniformly from $[-1,1]^2$ and recenter the coordinate system with that as the origin.
%
%algorithm: let $(x,y)$ be the lower vertex of the input square. consider the point $(x \bmod 2, y \bmod 2)$. if this point is contained in the region $[0,1)^2$, then select it as long as it doesn't intersect any already selected squares.
%
%let $\mathcal{D}$ be the input set. let $\mathcal{D}'$ be the squares that passed the algorithm's ``mod check''. the probability for each square of being in $\mathcal{D}'$ is $1/4$. let $A(\cdot)$ be the output of an optimal offline algorithm on a given input. let $B(\cdot)$ be the output of the online algorithm on a given input. we have $\E[|A(\mathcal{D}')|] \geq \E[|A(\mathcal{D}) \cap \mathcal{D}'|] = 1/4 \cdot |A(\mathcal{D})|$. note that the intersection graph of $\mathcal{D}'$ consists of disjoint cliques. in each clique, both the online algorithm and the optimal offline algorithm select exactly one square. therefore, $\E[|B(\mathcal{D}')|] = \E[|A(\mathcal{D}')|]$. this gives us an upper bound of 4 on the competitive ratio.
%
%this argument can be extended to $d$ dimensions to get a $2^d$ upper bound.

\begin{theorem}
    Every online algorithm must have competitive ratio at least $\frac{32}{15}$ asymptotically.
\end{theorem}

\begin{proof}
	The adversary begins in State $i$. The adversary first gives two disjoint unit hypercubes as input. Then, it (uniformly) randomly ``marks'' one (denote it by $x_{i,1}$) and gives as input two disjoint, unit hypercubes that intersect $x_{i,1}$ but no other hypercubes from the existing input. Next, it randomly marks one of these (denote it by $x_{i,2}$) and gives as input two disjoint, unit hypercubes that intersect $x_{i,1}$ and $x_{i,2}$ but no others. It continues this process until it has given 6 hypercubes, at the $j$th level marking a hypercube $x_{i,j}$ and giving as input two hypercubes that intersect $x_{i,1}$ through $x_{i,j}$ but no others. Once 6 hypercubes have been given in State $i$, the adversary switches to State $i+1$ and restarts the process from $j=1$. Once $n$ hypercubes have been given as input (by the end of State $I \coloneqq \lceil n/6 \rceil$), there is an MIS of size $\bigl\lfloor \frac{n}{2} \bigr\rfloor + I$ consisting of all the hypercubes not marked by the adversary.
	
	Consider any online algorithm. For each state $i$ and level $j$, by symmetry, the probability that it selects only $x_{i,j}$ and the probability that it selects only the other hypercube are equal; denote this probability by $p_{i,j}$. Let $p_{\sup} = \sup \set{p_{i,j}}$. Then, in each state $i \ne I$, the expected size of the independent set $X_i$ constructed by the online algorithm can be bounded as follows.
	\begin{align*}
		\E[X_i]
		&\leq 1 + p_{i,1} (1 + p_{i,2} (1 + p_{i,3})) \\
		&\leq 1 + p_{\sup} (1 + p_{\sup} (1 + p_{\sup})) \\
		&\leq 1 + \frac{1}{2} \biggl(1 + \frac{1}{2} \biggl(1 + \frac{1}{2}\biggr)\biggr) \\
		&= \frac{15}{8}
	\end{align*}
	The expected size of the overall independent set $\bigcup \set{X_i}_{i=1}^{I}$ is then at most $\frac{15}{8} I = \frac{15}{8} \bigl\lceil \frac{n}{6} \bigr\rceil$.
	Finally, the competitive ratio is at least
	\[
	\frac{\bigl\lfloor \frac{n}{2} \bigr\rfloor + \lceil \frac{n}{6} \rceil}{\frac{15}{8} \bigl\lceil \frac{n}{6} \bigr\rceil}
	= \frac{\bigl\lfloor \frac{n}{2} \bigr\rfloor}{\frac{15}{8} \bigl\lceil \frac{n}{6} \bigr\rceil} + \frac{8}{15}
	\,,\]
	and is asymptotically at least $\frac{32}{15}$.
\end{proof}

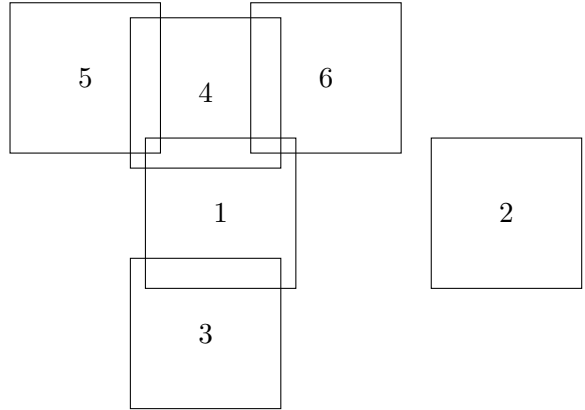
\begin{figure}[bhpt]
\centering
\begin{tikzpicture}[scale=2]
	\draw (2.5,2.5) rectangle node {1} (3.5,3.5);
	\draw (4.4,2.5) rectangle node {2} (5.4,3.5);
	\draw (2.4,1.7) rectangle node {3} (3.4,2.7);
	\draw (2.4,3.3) rectangle node {4} (3.4,4.3);
	\draw (1.6,3.4) rectangle node {5} (2.6,4.4);
	\draw (3.2,3.4) rectangle node {6} (4.2,4.4);
\end{tikzpicture}
\caption{An example of an oblivious adversary strategy for unit hypercubes in arbitrary order.}
\label{fig:adversary_strategy_unit_2}
\end{figure}

An example of the adversary strategy from the proof is shown in Figure~\ref{fig:adversary_strategy_unit_2}.

\section{\texorpdfstring{$\bm{\sigma}$}{σ}-Bounded Hypercubes}
\label{sec:sig-bounded}
For $\sigma \geq 1$, let a $\sigma$-bounded hypercube be a hypercube with side length between 1 and $\sigma$, inclusive. We assume that the online algorithm knows the value of $\sigma$. The results for the oblivious adversary model in this section use techniques based on those of \citet{Caragiannis2007}.

\subsection{Non-Dominated Order and Adaptive Adversaries}
\begin{theorem}
    \detgreedy{} has a competitive ratio of at most $(\lceil\sigma\rceil + 1)^d - \lceil\sigma\rceil^d$ under the adaptive offline adversary model.
\end{theorem}

\begin{proof}
	Let $A$ be the set of hypercubes in some optimal solution. Let $B$ be the set of hypercubes in the greedy solution. For each hypercube $x$ in $B \setminus A$, there are at most $(\lceil\sigma\rceil + 1)^d - \lceil\sigma\rceil^d$ hypercubes in $A \setminus B$ that intersect it and are not dominated by it. Every hypercube in $A \setminus B$ that intersects and is dominated by $x$ must intersect (and not be dominated by) some other hypercube in $B$ (or it would have been chosen by \detgreedy{}). Consequently, $A \setminus B$ can be partitioned into subsets of size at most $\bigl((\lceil\sigma\rceil + 1)^d - \lceil\sigma\rceil^d\bigr)$, each subset associated with a distinct hypercube in $B \setminus A$. Thus, we have the following.
	\begin{align*}
		|A \setminus B| &\leq \bigl((\lceil\sigma\rceil + 1)^d - \lceil\sigma\rceil^d\bigr) \cdot |B \setminus A| \\
		|A \setminus B| + |A \cap B| &\leq \bigl((\lceil\sigma\rceil + 1)^d - \lceil\sigma\rceil^d\bigr) \cdot |B \setminus A| + |A \cap B| \\
		|A \setminus B| + |A \cap B| &\leq \bigl((\lceil\sigma\rceil + 1)^d - \lceil\sigma\rceil^d\bigr) \cdot |B \setminus A| + \bigl((\lceil\sigma\rceil + 1)^d - \lceil\sigma\rceil^d\bigr) \cdot |A \cap B| \\
		|A \setminus B| + |A \cap B| &\leq \bigl((\lceil\sigma\rceil + 1)^d - \lceil\sigma\rceil^d\bigr) \cdot ( |B \setminus A| + |A \cap B| ) \\
		|A| &\leq \bigl((\lceil\sigma\rceil + 1)^d - \lceil\sigma\rceil^d\bigr) \cdot |B|
	\end{align*}
\end{proof}

\begin{theorem}\label{thm:sig-bounded:non-dominated:adaptive-online}
    Every online algorithm must have competitive ratio at least $(\lceil\sigma\rceil + 1)^d - \lceil\sigma\rceil^d$ under the adaptive online adversary model.
\end{theorem}

\begin{proof}
	Consider any algorithm. Until the algorithm selects a hypercube, the adversary gives disjoint hypercubes with side length $\sigma$. Once the algorithm selects one, the adversary gives as input $(\lceil\sigma\rceil + 1)^d - \lceil\sigma\rceil^d$ disjoint unit hypercubes intersecting that one.
\end{proof}

\begin{corollary}
    \detgreedy{} is optimal for $\sigma$-bounded hypercubes in non-dominated order under adaptive adversary models.
\end{corollary}

\subsection{Non-Dominated Order and an Oblivious Adversary}
\label{sec:sig-bounded:non-dominated:oblivious}
For any integer $k \geq 1$, let $b = \sigma^{1/k}$.
Consider the following algorithm. First, it picks an integer $i$ uniformly at random from the interval $[0, k-1]$.
Then it greedily selects each hypercube with side length in the range $[b^i, b^{i+1}]$ and ignores the rest. Denote this algorithm by \selgreedy{k}.

\begin{theorem}
    \selgreedy{k} has a competitive ratio of $(\lceil b \rceil + 1)^d k - \lceil b \rceil^d k$.
\end{theorem}

\begin{proof}
	Let $A$ be the MIS constructed by an optimal offline algorithm, $B$ be the independent set constructed by the online algorithm, $B'$ be the set of hypercubes in the input with side length in the range $[b^i, b^{i+1}]$ (those not ignored by the online algorithm), and $A'$ be a MIS of the hypercubes in $B'$.
	
	If we consider only $B'$, then the input consists of (potentially scaled-up) $b$-bounded hypercubes. We know that the greedy algorithm has a competitive ratio of $(\lceil b \rceil + 1)^d - \lceil b \rceil^d$ in this case, so the size of $A'$ is at most $((\lceil b \rceil + 1)^d - \lceil b \rceil^d) \cdot |B|$. The size of $A'$ is necessarily at least $|A \cap B'|$ since $A'$ is a MIS of the hypercubes in $B'$, and the expected size of $A \cap B'$ is $\frac{1}{k} |A|$. Thus, the competitive ratio of the full online algorithm is at most $(\lceil b \rceil + 1)^d k - \lceil b \rceil^d k$.
\end{proof}

The next question is how to choose the value of $k$. Notably, if we take $k=1$, we have the deterministic online algorithm used for the adaptive adversary case and the same competitive ratio of $(\lceil\sigma\rceil + 1)^d - \lceil\sigma\rceil^d$. At the other extreme, assuming $\sigma \geq 2$, if we take $k = \lceil \log_2 \sigma \rceil$, we get $\lceil b \rceil = 2$ and a competitive ratio of at most $3^d \lceil \log_2 \sigma \rceil - 2^d \lceil \log_2 \sigma \rceil$. For sufficiently large $d$, the latter is a significantly better bound. The optimal $k$ is somewhere in the interval $[1, \lceil \log_2 \sigma \rceil]$; determining the exact optimum is an open problem.

\begin{theorem}
    Every online algorithm must have competitive ratio at least $\frac{\lceil \log_2 \sigma \rceil + 1}{2}$ asymptotically.
\end{theorem}

\begin{proof}
	The adversary begins in State $i$. The adversary first gives two disjoint hypercubes as input. Then, it (uniformly) randomly ``marks'' one (denote it by $x_{i,1}$) and gives as input two disjoint hypercubes that intersect $x_{i,1}$ but no other hypercubes from the existing input. Next, it randomly marks one of these (denote it by $x_{i,2}$) and gives as input two disjoint hypercubes that intersect $x_{i,1}$ and $x_{i,2}$ but no others. It continues this process until it has given $2\lceil \log_2 \sigma \rceil$ hypercubes, at the $j$th level marking a hypercube $x_{i,j}$ and giving as input two hypercubes that intersect $x_{i,1}$ through $x_{i,j}$ but no others. Once $2\lceil \log_2 \sigma \rceil$ hypercubes have been given in State $i$, the adversary switches to State $i+1$ and restarts the process from $j=1$. Once $n$ hypercubes have been given as input (by the end of State $I \coloneqq \lceil n/(2\lceil \log_2 \sigma \rceil) \rceil$), there is an MIS of size $\bigl\lfloor \frac{n}{2} \bigr\rfloor + I$ consisting of all the hypercubes not marked by the adversary.
	
	Consider any online algorithm. For each state $i$ and level $j$, by symmetry, the probability that it selects only $x_{i,j}$ and the probability that it selects only the other hypercube are equal; denote this probability by $p_{i,j}$. Let $p_{\sup} = \sup \set{p_{i,j}}$. Then, in each state $i$, the expected size of the independent set $X_i$ constructed by the online algorithm can be bounded as follows.
	\begin{align*}
		\E[X_i]
		&\leq 1 + p_{i,1} (1 + p_{i,2} (1 + \ldots + p_{i,\lceil \log_2 \sigma \rceil})\dots)) \\
		&\leq 1 + p_{i,1} (1 + p_{i,2} (1 + \dots)) \\
		&\leq 1 + p_{\sup} (1 + p_{\sup} (1 + \dots)) \\
		&= \frac{1}{1 - p_{\sup}} \\
		&\leq \frac{1}{1 - \frac{1}{2}} \\
		&= 2
	\end{align*}
	The expected size of the overall independent set $\bigcup \set{X_i}_{i=1}^{I}$ is then at most $2I = 2\lceil n/(2\lceil \log_2 \sigma \rceil) \rceil$.
	Finally, the competitive ratio is at least
	\[
	\frac{\bigl\lfloor \frac{n}{2} \bigr\rfloor + \lceil n/(2\lceil \log_2 \sigma \rceil) \rceil}{2\lceil n/(2\lceil \log_2 \sigma \rceil) \rceil}
	= \frac{\bigl\lfloor \frac{n}{2} \bigr\rfloor}{2\lceil n/(2\lceil \log_2 \sigma \rceil) \rceil} + \frac{1}{2}
	\,,\]
	which is asymptotically at least $\frac{\lceil \log_2 \sigma \rceil + 1}{2}$.
\end{proof}

\subsection{Arbitrary Order and Adaptive Adversaries}
\begin{theorem}
    \detgreedy{} has a competitive ratio of at most $(\lceil\sigma\rceil + 1)^d$ under the adaptive offline adversary model.
\end{theorem}

\begin{proof}
	Note that a $\sigma$-bounded $d$-cube can be intersected by at most $(\lceil\sigma\rceil + 1)^d$ disjoint $\sigma$-bounded $d$-cubes. Let $A$ be the set of hypercubes in some optimal solution. Let $B$ be the set of hypercubes in the greedy solution. For each hypercube in $B \setminus A$, there are at most $(\lceil\sigma\rceil + 1)^d$ hypercubes in $A \setminus B$ that intersect it. By construction, every hypercube in $A \setminus B$ must intersect some hypercube in $B \setminus A$ (or it would be in $B$). Consequently, $A \setminus B$ can be partitioned into subsets of size at most $(\lceil\sigma\rceil + 1)^d$, each subset associated with a distinct hypercube in $B \setminus A$. Thus, we have the following.
	\begin{align*}
		|A \setminus B| &\leq (\lceil\sigma\rceil + 1)^d \cdot |B \setminus A| \\
		|A \setminus B| + |A \cap B| &\leq (\lceil\sigma\rceil + 1)^d \cdot |B \setminus A| + |A \cap B| \\
		|A \setminus B| + |A \cap B| &\leq (\lceil\sigma\rceil + 1)^d \cdot |B \setminus A| + (\lceil\sigma\rceil + 1)^d \cdot |A \cap B| \\
		|A \setminus B| + |A \cap B| &\leq (\lceil\sigma\rceil + 1)^d \cdot ( |B \setminus A| + |A \cap B| ) \\
		|A| &\leq (\lceil\sigma\rceil + 1)^d \cdot |B|
	\end{align*}
\end{proof}

\begin{theorem}
    Every online algorithm must have competitive ratio at least $(\lceil\sigma\rceil + 1)^d$ under the adaptive online adversary model.
\end{theorem}

\begin{proof}
	Consider any algorithm. Until the algorithm selects a hypercube, the adversary gives disjoint hypercubes with side length $\sigma$. Once the algorithm selects one, the adversary gives as input $(\lceil\sigma\rceil + 1)^d$ disjoint unit hypercubes intersecting that one.
\end{proof}

\begin{corollary}
    \detgreedy{} is optimal for $\sigma$-bounded hypercubes in arbitrary order under adaptive adversary models.
\end{corollary}

\subsection{Arbitrary Order and an Oblivious Adversary}
\label{sec:sig-bounded:arbitrary:oblivious}
\begin{theorem}
    \selgreedy{k} has a competitive ratio of $(\lceil b \rceil + 1)^d k$.
\end{theorem}

\begin{proof}
	Let $A$ be the MIS constructed by an optimal offline algorithm, $B$ be the independent set constructed by the online algorithm, $B'$ be the set of hypercubes in the input with side length in the range $[b^i, b^{i+1}]$ (those not ignored by the online algorithm), and $A'$ be a MIS of the hypercubes in $B'$.
	
	If we consider only $B'$, then the input consists of (potentially scaled-up) $b$-bounded hypercubes. We know that the greedy algorithm has a competitive ratio of $(\lceil b \rceil + 1)^d$ in this case, so the size of $A'$ is at most $(\lceil b \rceil + 1)^d \cdot |B|$. The size of $A'$ is necessarily at least $|A \cap B'|$ since $A'$ is a MIS of the hypercubes in $B'$, and the expected size of $A \cap B'$ is $\frac{1}{k} |A|$. Thus, the competitive ratio of the full online algorithm is at most $(\lceil b \rceil + 1)^d k$.
\end{proof}

The next question is how to choose the value of $k$. Notably, if we take $k=1$, we have the deterministic online algorithm used for the adaptive adversary case and the same competitive ratio of $(\lceil\sigma\rceil + 1)^d$. At the other extreme, assuming $\sigma \geq 2$, if we take $k = \lceil \log_2 \sigma \rceil$, we get $\lceil b \rceil = 2$ and a competitive ratio of at most $3^d \lceil \log_2 \sigma \rceil$.
The optimal $k$ is somewhere in the interval $[1, \lceil \log_2 \sigma \rceil]$.

To find a good choice of $k$, we differentiate the upper bound on the competitive ratio with respect to $k$.
For simpler analysis, we use a looser form of the upper bound: $(b + 2)^d k$. We have the following.
\begin{align*}
	\frac{d}{dk} \bigl(\sigma^{1/k} + 2\bigr)^d k
	&= \bigl(\sigma^{1/k} + 2\bigr)^d + d \bigl(\sigma^{1/k} + 2\bigr)^{d-1} \ln(\sigma) \cdot \sigma^{1/k} \bigl(-k^{-2}\bigr) k \\
	&= \bigl(\sigma^{1/k} + 2\bigr)^d - d \bigl(\sigma^{1/k} + 2\bigr)^{d-1} \ln(\sigma) \cdot \sigma^{1/k} k^{-1} \\
	&= \bigl(\sigma^{1/k} + 2\bigr)^{d-1} \bigl(\sigma^{1/k} + 2 - d \ln(\sigma) \cdot \sigma^{1/k} k^{-1}\bigr) \\
	&= \bigl(\sigma^{1/k} + 2\bigr)^{d-1} \bigl(2 + \sigma^{1/k} \bigl(1 - d \ln(\sigma) \cdot k^{-1}\bigr)\bigr)
\end{align*}
For very small $k$, the derivative is negative, at $k = d \ln \sigma$, it is positive, and in between, it changes sign exactly once.

\begin{claim}
    The derivative of $(b + 2)^d k$ with respect to $k$ is $0$ at
    \[k' = \frac{\ln \sigma}{W_0\Bigl(\frac{2e^{-1/d}}{d}\Bigr) + \frac{1}{d}} \,,\]
    where $W_0$ is the principal branch of the Lambert $W$ function.
\end{claim}

\begin{proof}
We verify that $k'$ is the solution to the equation by substitution.
\begin{align*}
	&\phanrel 2 + \sigma^{1/k'} \bigl(1 - d \ln(\sigma) \cdot (k')^{-1}\bigr) \\
	&= 2 + \sigma^{1/k'} \Biggl(1 - d \ln(\sigma) \cdot \frac{W_0\Bigl(\frac{2e^{-1/d}}{d}\Bigr) + \frac{1}{d}}{\ln \sigma}\Biggr) \\
	&= 2 - d\sigma^{1/k'} W_0\biggl(\frac{2e^{-1/d}}{d}\biggr) \\
	&= 2 - d\sigma^{\frac{W_0\bigl(\frac{2e^{-1/d}}{d}\bigr) + \frac{1}{d}}{\ln \sigma}} W_0\biggl(\frac{2e^{-1/d}}{d}\biggr) \\
	&= 2 - de^{W_0\bigl(\frac{2e^{-1/d}}{d}\bigr) + \frac{1}{d}} W_0\biggl(\frac{2e^{-1/d}}{d}\biggr) \\
	&= 2 - d e^{1/d} e^{W_0\bigl(\frac{2e^{-1/d}}{d}\bigr)} W_0\biggl(\frac{2e^{-1/d}}{d}\biggr) \\
	&= 2 - d e^{1/d} \frac{2e^{-1/d}}{d} \\
	&= 2 - 2e^0 \\
	&= 0
\end{align*}
\end{proof}

We now have the zero of the derivative, but without evaluating the Lambert $W$ function, the expression is difficult to interpret. Accordingly, we provide approximate values of $\Bigl(W_0\Bigl(\frac{2e^{-1/d}}{d}\Bigr) + \frac{1}{d}\Bigr)^{-1}$ in various dimensions $d$ below.
\begin{center}
\begin{tabular}{cccccccc}
    \toprule
    $d$ & 1 & 2 & 3 & 4 & 10 & 100 & 1000 \\
    \midrule
    value & 0.683501 & 1.10537 & 1.48514 & 1.84821 & 3.92179 & 33.9903 & 333.999 \\
    \bottomrule
\end{tabular}
\end{center}
For very small $x$, we have $W_0(x) \approx x$; thus, for large $d$, we have
\begin{equation*}
	\biggl(W_0\biggl(\frac{2e^{-1/d}}{d}\biggr) + \frac{1}{d}\biggr)^{-1}
	\approx \biggl(W_0\biggl(\frac{2}{d}\biggr) + \frac{1}{d}\biggr)^{-1}
	\approx \biggl(\frac{2}{d} + \frac{1}{d}\biggr)^{-1}
	= \frac{d}{3} \,,
\end{equation*}
which gives $k' \approx \frac{d}{3} \ln \sigma$. For large $d$, this is greater than $\lceil \log_2 \sigma \rceil$, so the optimal $k$ is likely $\lceil \log_2 \sigma \rceil$.
For small $d$, the easiest approach is to evaluate $(\lceil b \rceil + 1)^d k$ for several integers $k$ near $k'$ and select the best one.

% Finally, the lower bound from \S~\ref{sec:sig-bounded:non-dominated:oblivious} applies in this setting as well.

\section{Unit-Volume Hyperrectangles}
\label{sec:unit-vol}
The results in this section apply to both non-dominated and arbitrary orders.

\begin{theorem}
    \detgreedy{} has a competitive ratio of at most $n - 1$ under the adaptive offline adversary model.
\end{theorem}

\begin{proof}
    As long as $n > 0$, \detgreedy{} will select at least one hyperrectangle.
	Suppose some optimal solution has $n$ hyperrectangles. Then, all $n$ hyperrectangles in the input are disjoint, so \detgreedy{} will also select $n$ hyperrectangles, giving a ratio of 1. If the optimal solution has at most $n - 1$ hyperrectangles, then the ratio is at most $n - 1$.
\end{proof}

\begin{theorem}
    Every online algorithm must have competitive ratio at least $n - 1$ under the adaptive online adversary model.
\end{theorem}

\begin{proof}
	The adversary can give as input disjoint, dominating, unit-volume hyperrectangles until the algorithm selects one. Then, the adversary can have the remaining input be disjoint, unit-volume hyperrectangles in dominating order that intersect (and are not dominated by) the selected hyperrectangle.
\end{proof}

\begin{figure}[bhpt]
\centering
\begin{tikzpicture}[scale=1]
	\draw (1.5,1.5) rectangle node [shift={(0,0.55)}] {1} (2.167,3.0);
	\draw (1.6,1.6) rectangle node [at end, shift={(0.2,-0.05)}] {2} (11.6,1.7);
	\draw (1.6,1.9) rectangle node [at end, shift={(0.2,-0.05)}] {3} (11.6,2.0);
	\draw (1.6,2.2) rectangle node [at end, shift={(0.2,-0.05)}] {4} (11.6,2.3);
	\draw (1.6,2.5) rectangle node [at end, shift={(0.2,-0.05)}] {5} (11.6,2.6);
\end{tikzpicture}
\caption{An example of an adaptive online adversary strategy for unit-volume hyperrectangles in non-dominated order.}
\label{fig:adversary_strategy_unit-vol_1}
\end{figure}
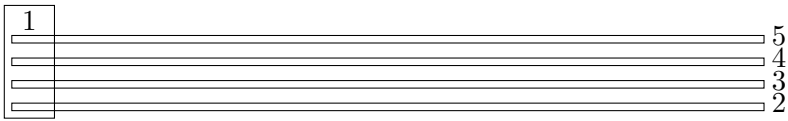

An example of the adversary strategy from the proof is shown in Figure~\ref{fig:adversary_strategy_unit-vol_1}.

\begin{corollary}
    \detgreedy{} is optimal for unit-volume hyperrectangles under adaptive adversary models.
\end{corollary}

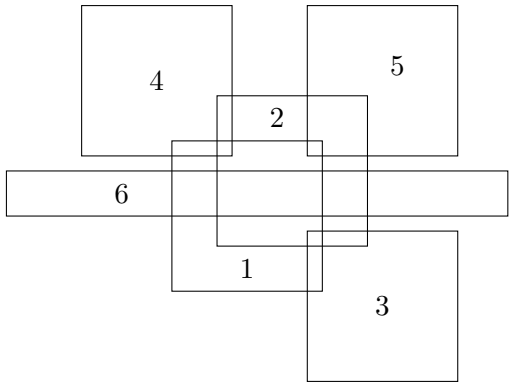
\begin{figure}[bhpt]
\centering
\begin{tikzpicture}[scale=2]
	\draw (-0.1,-0.1) rectangle node [shift={(0.0,-0.7)}] {1} (0.9,0.9);
	\draw (0.2,0.2) rectangle node [shift={(-0.2,0.7)}] {2} (1.2,1.2);
	\draw (1.8,-0.7) rectangle node {3} (0.8,0.3);
	\draw (-0.7,1.8) rectangle node {4} (0.3,0.8);
	\draw (1.8,1.8) rectangle node [shift={(0.2,0.2)}] {5} (0.8,0.8);
	\draw (-1.2,0.4) rectangle node [at start, shift={(-0.2,0.3)}] {6} (2.133,0.7);
\end{tikzpicture}
\caption{An optimal arrangement of $n=6$ unit-area rectangles for $p=0.50$.}
\label{fig:n6_unit-vol}
\end{figure}

Next, under the oblivious adversary model, all results on \greedy{p} for fixed $n$ from \S~\ref{sec:unit:non-dominated:oblivious} (unit hypercubes) hold here. In addition, we can construct an arrangement for $n=6$ (see Figure~\ref{fig:n6_unit-vol}), so the result for $n=6$ from \S~\ref{sec:unit:arbitrary:oblivious} holds as well.

\begin{theorem}\label{thm:unit-vol:non-dominated:oblivious}
    Every online algorithm must have competitive ratio at least $\frac{1}{2} \bigl\lfloor \frac{n}{2} \bigr\rfloor + \frac{1}{2}$.
\end{theorem}

\begin{proof}
	The adversary first gives two disjoint, unit-volume hyperrectangles as input. Then, it (uniformly) randomly ``marks'' one (denote it by $x_1$) and gives as input two disjoint hyperrectangles that intersect $x_1$ but no other hyperrectangles from the existing input. Next, it randomly marks one of these (denote it by $x_2$) and gives as input two disjoint hyperrectangles that intersect $x_1$ and $x_2$ but no others. It continues this process indefinitely, at the $k$th level marking a hyperrectangle $x_k$ and giving as input two hyperrectangles that intersect $x_1$ through $x_k$ but no others. Once $n$ hyperrectangles have been given as input, there is an MIS of size $\bigl\lfloor \frac{n}{2} \bigr\rfloor + 1$ consisting of all the hyperrectangles not marked by the adversary.
	
	Consider any online algorithm. At each level $k$, by symmetry, the probability that it selects only $x_k$ and the probability that it selects only the other hyperrectangle are equal; denote this probability by $p_k$. Let $p_{\sup} = \sup \set{p_k}$. Then, the expected size of the independent set $X$ constructed by the online algorithm can be bounded as follows.
	\begin{align*}
		\E[X]
		&\leq 1 + p_1 (1 + p_2 (1 + \dots)) \\
		&\leq 1 + p_{\sup} (1 + p_{\sup} (1 + \dots)) \\
		&= \frac{1}{1 - p_{\sup}} \\
		&\leq \frac{1}{1 - \frac{1}{2}} \\
		&= 2
	\end{align*}
	The competitive ratio is then at least
	$\frac{1}{2} \bigl\lfloor \frac{n}{2} \bigr\rfloor + \frac{1}{2}$,
	which is $\Theta(n)$.
\end{proof}

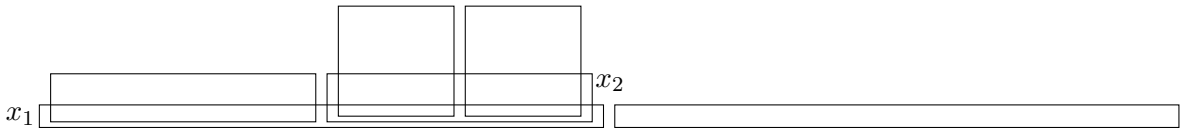
\begin{figure}[bhpt]
\centering
\begin{tikzpicture}[scale=1.5]
	\draw (0,0) rectangle node [shift={(-4,0)}] {$x_1$} (5,0.2);
	% \draw (5.1,0) rectangle (10.1,0.2);
    \draw (8.5,0) -- (5.1,0) -- (5.1,0.2) -- (8.5,0.2);
	\draw (0.1,0.05) rectangle (2.45,0.476);
	\draw (2.55,0.05) rectangle node [shift={(2,0.2)}] {$x_2$} (4.9,0.476);
	\draw (2.65,0.1) rectangle (3.675,1.076);
	\draw (3.775,0.1) rectangle (4.8,1.076);
\end{tikzpicture}
\caption{An example of an oblivious adversary strategy for unit-volume hyperrectangles in non-dominated order.}
\label{fig:adversary_strategy_unit-vol_2}
\end{figure}

An example illustrating the adversary strategy from the proof is shown in Figure~\ref{fig:adversary_strategy_unit-vol_2}.

\section{Arbitrary Hypercubes}
\label{sec:cube}
The results in this section apply to both non-dominated and arbitrary orders.

\begin{theorem}
    \detgreedy{} has a competitive ratio of at most $n - 1$ under the adaptive offline adversary model.
\end{theorem}

\begin{proof}
    As long as $n > 0$, \detgreedy{} will select at least one hypercube.
	Suppose some optimal solution has $n$ hypercubes. Then, all $n$ hypercubes in the input are disjoint, so \detgreedy{} will also select $n$ hypercubes, giving a ratio of 1. If the optimal solution has at most $n - 1$ hypercubes, then the ratio is at most $n - 1$.
\end{proof}

\begin{theorem}
    Every online algorithm must have competitive ratio at least $n - 1$ under the adaptive online adversary model.
\end{theorem}

\begin{proof}
	The adversary can give as input disjoint, dominating hypercubes until the algorithm selects one. Then, the adversary can have the remaining input be disjoint hypercubes in non-dominated order that intersect (and are not dominated by) the selected hypercube.
\end{proof}

\begin{figure}[bhpt]
\centering
\begin{tikzpicture}[scale=1]
	\draw (1.5,1.5) rectangle node {1} (3.5,3.5);
	\draw (3.4,1.6) rectangle node [shift={(0.3,0)}] {2} (3.6,1.8);
	\draw (3.4,1.9) rectangle node [shift={(0.3,0)}] {3} (3.6,2.1);
	\draw (3.4,2.2) rectangle node [shift={(0.3,0)}] {4} (3.6,2.4);
	\draw (3.4,2.5) rectangle node [shift={(0.3,0)}] {5} (3.6,2.7);
\end{tikzpicture}
\caption{An example of an adaptive online adversary strategy for hypercubes in non-dominated order.}
\label{fig:adversary_strategy_cube_1}
\end{figure}
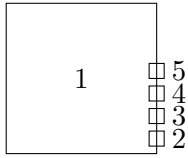

An example of the adversary strategy from the proof is shown in Figure~\ref{fig:adversary_strategy_cube_1}.

\begin{corollary}
    \detgreedy{} is optimal for arbitrary hypercubes under adaptive adversary models.
\end{corollary}

\begin{figure}[bhpt]
\centering
\begin{tikzpicture}[scale=1.5]
	\draw (-0.1,-0.1) rectangle node [shift={(-0.4,-1.25)}] {1} (1.9,1.9);
	\draw (0.2,0.2) rectangle node [shift={(0.4,1.25)}] {2} (2.2,2.2);
	\draw (2.8,-0.7) rectangle node {3} (1.8,0.3);
	\draw (-0.7,2.8) rectangle node {4} (0.3,1.8);
    \draw (1.8,0.55) rectangle node [shift={(0.2,0.0)}] {5} (2.8,1.55);
	\draw (2.8,2.8) rectangle node [shift={(0.2,0.2)}] {6} (1.8,1.8);
\end{tikzpicture}
\caption{An optimal arrangement of $n=6$ squares for $p=0.50$.}
\label{fig:n6_cube}
\end{figure}
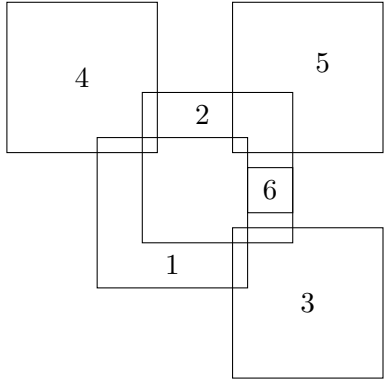

Next, under the oblivious adversary model, all results on \greedy{p} for fixed $n$ from \S~\ref{sec:unit:non-dominated:oblivious} (unit hypercubes) hold here. In addition, we can construct an arrangement for $n=6$ (see Figure~\ref{fig:n6_cube}), so the result for $n=6$ from \S~\ref{sec:unit:arbitrary:oblivious} holds as well.

\begin{theorem}
    Every online algorithm must have competitive ratio at least $\frac{1}{2} \bigl\lfloor \frac{n}{2} \bigr\rfloor + \frac{1}{2}$.
\end{theorem}

\begin{proof}
	Suppose $n$ is finite.
	The adversary first gives two disjoint hypercubes as input. Then, it (uniformly) randomly ``marks'' one (denote it by $x_1$) and gives as input two disjoint hypercubes that intersect $x_1$ but no other hypercubes from the existing input. Next, it randomly marks one of these (denote it by $x_2$) and gives as input two disjoint hypercubes that intersect $x_1$ and $x_2$ but no others. It continues this process until it has given $n$ hypercubes as input, at the $k$th level marking a hypercube $x_k$ and giving as input two hypercubes that intersect $x_1$ through $x_k$ but no others. At the end of this process, there is an MIS of size $\bigl\lfloor \frac{n}{2} \bigr\rfloor + 1$ consisting of all the hypercubes not marked by the adversary.
	
	Consider any online algorithm. At each level $k$, by symmetry, the probability that it selects only $x_k$ and the probability that it selects only the other hypercube are equal; denote this probability by $p_k$. Let $p_{\sup} = \sup \set{p_k}$. Then, the expected size of the independent set $X$ constructed by the online algorithm can be bounded as follows.
	\begin{align*}
		\E[X]
		&\leq 1 + p_1 (1 + p_2 (1 + \dots)) \\
		&\leq 1 + p_{\sup} (1 + p_{\sup} (1 + \dots)) \\
		&= \frac{1}{1 - p_{\sup}} \\
		&\leq \frac{1}{1 - \frac{1}{2}} \\
		&= 2
	\end{align*}
	The competitive ratio is then at least
	$\frac{1}{2} \bigl\lfloor \frac{n}{2} \bigr\rfloor + \frac{1}{2}$,
	which is $\Theta(n)$.
	
	If $n = \infty$, then for any claimed finite competitive ratio $\alpha$, the adversary can repeatedly execute the above process with $2\alpha$ full levels for a competitive ratio of at least $\alpha + \frac{1}{2}$. Thus, the competitive ratio must be $\infty$.
\end{proof}

\begin{figure}[bhpt]
\centering
\begin{tikzpicture}[scale=1]
	\draw (0,0) rectangle node [shift={(0,-0.5)}] {$x_1$} (5,5);
	\draw (5.1,0) rectangle (10.1,5);
	\draw (0.1,2.75) rectangle (2.45,5.1);
	\draw (2.55,2.75) rectangle node [shift={(0,-0.235)}] {$x_2$} (4.9,5.1);
	\draw (2.65,4.175) rectangle (3.675,5.2);
	\draw (3.775,4.175) rectangle (4.8,5.2);
\end{tikzpicture}
\caption{An example of an oblivious adversary strategy for hypercubes in non-dominated order.}
\label{fig:adversary_strategy_cube_2}
\end{figure}
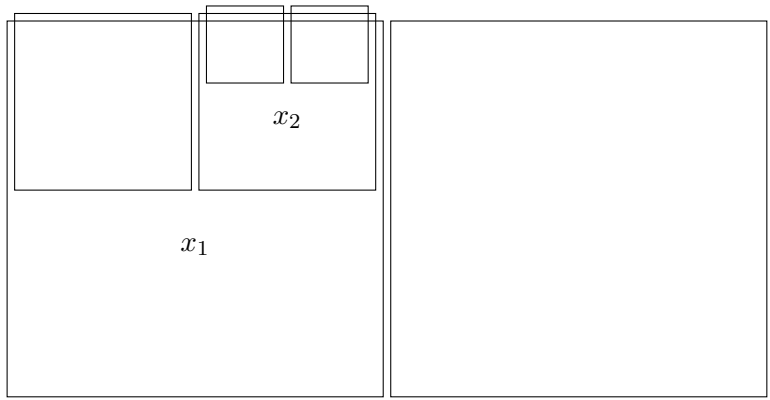

An example illustrating the adversary strategy from the proof is shown in Figure~\ref{fig:adversary_strategy_cube_2}.

\section{Future Work}
\label{sec:future_work}
In addition to the open problems mentioned throughout the paper, there are many more directions for future work.
First, there are several open questions surrounding the \greedy{p} algorithm under the oblivious adversary model:
\begin{itemize}
	\item Is it always optimal for the adversary to give the nodes with highest degree first?
	\item Does the fixed-$n$ competitive ratio (for optimal $p$) increase monotonically with $n$?
    \item What are the fixed-$n$ competitive ratios when $d=1$?
    \item Can we prove an upper bound on the general competitive ratio?
\end{itemize}

Next, for our bounds on the competitive ratio for $\sigma$-bounded hypercubes, we assume that the online algorithm has access to the value of $\sigma$. How would the results change if the algorithm does not know $\sigma$ in advance?
We also assume that the arrangement of hyperrectangles in space is given to the algorithm. It would be interesting to see which results still hold if only the intersection graph is revealed to the algorithm.

Another direction for future work would be to consider dominating/non-dominated order with respect to other points in the hyperrectangles (e.g., the centers or lower vertices).
Other input orders could be considered as well, including ordering by distances of the centers or lower/upper vertices of the hyperrectangles from the origin.

Finally, we could require that all hyperrectangles have integer coordinates. This could be combined with a requirement that all hyperrectangles in the input are contained within the region $[0,c]^d$ for some constant $c \geq 2$.
We could also consider a model where, instead of only being able to accept hyperrectangles from the input, the algorithm could at each time step choose to either accept a hyperrectangle or discard a previously selected one.

\section*{Acknowledgments}
This work was supported in part by the National Science Foundation (Grant No.~2107290) and the UIC University Fellowship.

\bibliographystyle{plainnat}
\bibliography{main}

\appendix

\section{Applications}
\label{sec:applications}
In this section, we give an overview of some of the applications of the online MIS of hyperrectangles problem.

\subsection{Range Queries}
Consider a database setting where a user needs to perform many expensive range queries. If we cache the results of a query, then we can use them to speed up execution of any range query that includes as a subset the earlier range. We need to ensure that we do not cache the results from overlapping ranges, as we do not want duplicates in the results. Furthermore, we need to decide which results to cache in an online manner, as we may not have enough space to store the results of all the queries. Since each range can be interpreted as a hyperrectangle, this is equivalent to the online MIS of hyperrectangles problem.

\subsection{Urban Planning}
For an example using non-dominated order, consider an urban planning setting where we have two dimensions: the location of buildings along a road and the height of those buildings. Each rectangle represents a potential residence, and a selected one is one that is actually built. If the view at the end of the road is spectacular and residents want to be able to see it from the rooftop, then each potential building must be either taller or closer to the end of the road than those proposed so far (i.e., it must be Pareto-efficient with respect to height and location).

If the two axes are instead latitude and longitude, then the problem of allocating land can be modeled as online MIS in the plane (using arbitrary order). Each rectangle represents a potential lot, and a selected one is one that is allocated. Potential buyers will request a piece of land, and the entity in charge must decide to grant the land to them or not. Whatever decision they make is irreversible (if they accept, they can't renege on the contract, and if they refuse, we assume the buyer will spend their money elsewhere).

\section{Experiments}
\label{sec:experiments}
\begin{figure}[bhpt]
	\centering
	\includegraphics[alt={Many overlapping rectangles, some shaded.}, width=0.7\linewidth]{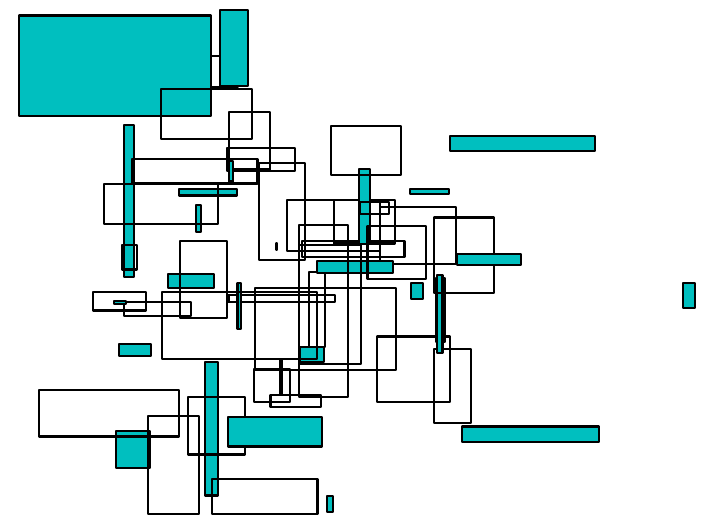}
	\caption{A visualization of a sample execution of the offline greedy algorithm. Rectangles selected by the algorithm are shaded.}
	\label{fig:vis}
\end{figure}

We ran several experiments to compare the performance of \detgreedy{} with an offline algorithm that greedily selects rectangles in order of increasing $x$ coordinates of upper vertices\footnote{All experiments were run on a machine with an Intel\textsuperscript{\textregistered{}} Core\texttrademark{} i5-8265U CPU and 16.0 GB of memory running Windows 11 Pro. All code is freely available online~\citep{Advani2024}.}. A visualization of the offline greedy algorithm can be seen in Figure~\ref{fig:vis}.

For each experiment, two distributions $\mathcal{D}_1$ and $\mathcal{D}_2$ are selected. $\mathcal{D}_1$ is chosen to be either the standard normal distribution or the uniform distribution on $[0,1)$. Given a location parameter $\ell$ and scale parameter $s$, $\mathcal{D}_2$ is chosen to be either the arcsine distribution with those parameters or a degenerate distribution that always returns $\ell + s$ (this distribution ends up forcing the rectangles to be squares with side length $s$). Each rectangle given as input is generated as follows. Two values are sampled from $\mathcal{D}_1$; these will be the coordinates of one vertex of the rectangle. Next, two values are sampled from $\mathcal{D}_2$ with $\ell$ set to the corresponding coordinate of the initial vertex; these will be the coordinates of the opposite vertex of the rectangle.

\begin{figure}[bhpt]
	\centering
	\includegraphics[alt={A plot of proportion of input selected going down and running time going up for two algorithms as input size increases.}, width=0.45\linewidth]{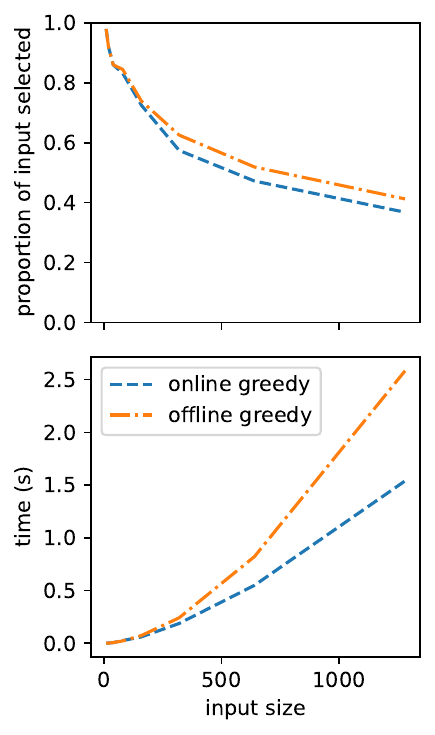}
    \includegraphics[alt={A plot of proportion of input selected going down and running time going down for two algorithms as scale parameter increases.}, width=0.45\linewidth]{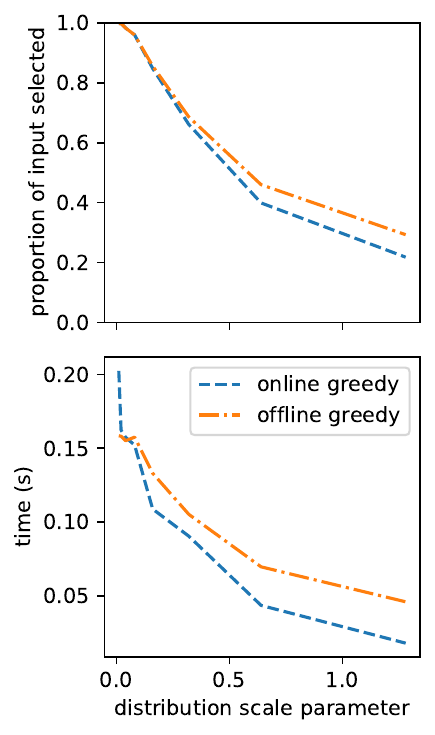}
	\caption{The performance of the online and offline greedy algorithms with $\mathcal{D}_1$ as the normal distribution and $\mathcal{D}_2$ as an arcsine distribution.}
	\label{fig:exp1}
\end{figure}

\begin{figure}[bhpt]
	\centering
	\includegraphics[alt={A plot of proportion of input selected going down and running time going up for two algorithms as input size increases.}, width=0.45\linewidth]{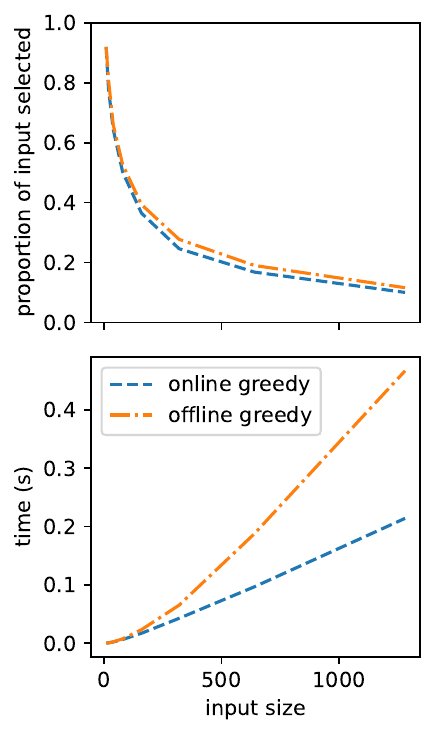}
    \includegraphics[alt={A plot of proportion of input selected going down and running time going down for two algorithms as scale parameter increases.}, width=0.45\linewidth]{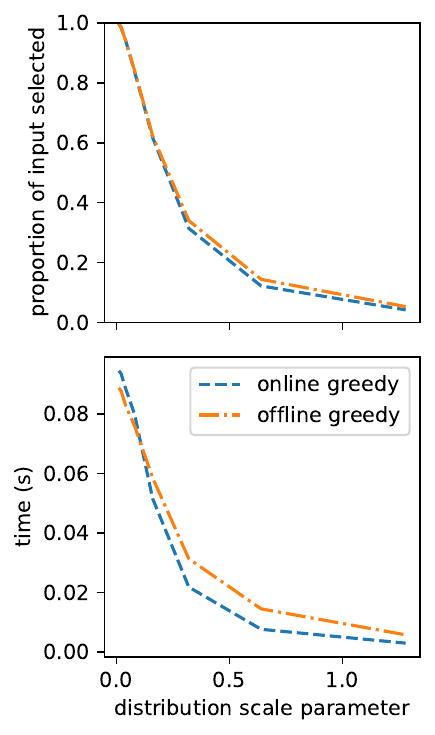}
	\caption{The performance of the online and offline greedy algorithms with $\mathcal{D}_1$ as the normal distribution and $\mathcal{D}_2$ as a degenerate distribution.}
	\label{fig:exp2}
\end{figure}

\begin{figure}[bhpt]
	\centering
	\includegraphics[alt={A plot of proportion of input selected going down and running time going up for two algorithms as input size increases.}, width=0.45\linewidth]{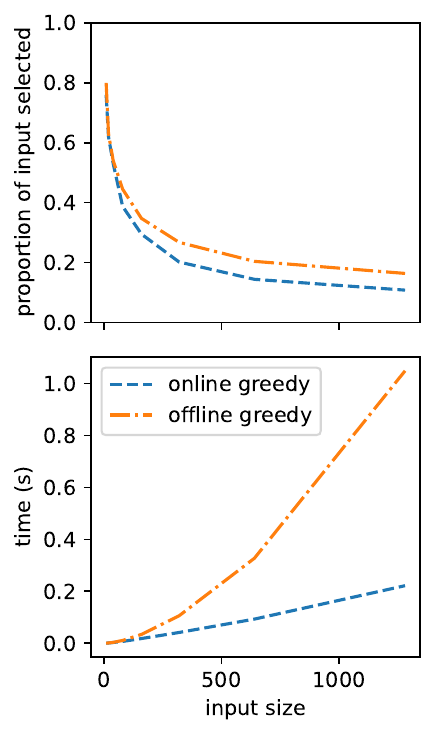}
    \includegraphics[alt={A plot of proportion of input selected going down and running time going down for two algorithms as scale parameter increases.}, width=0.45\linewidth]{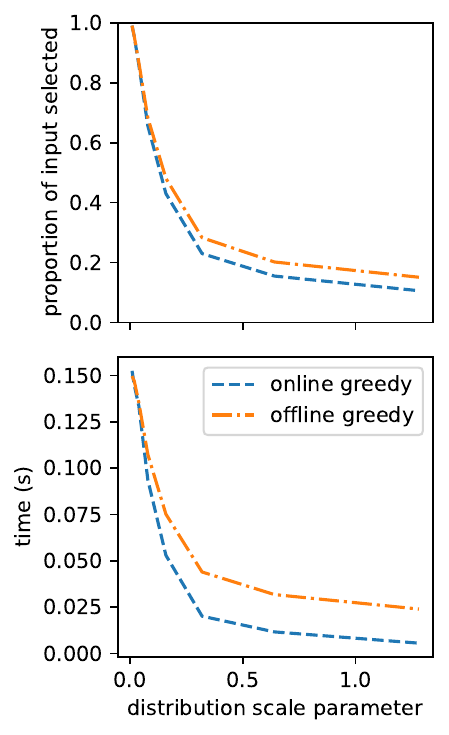}
	\caption{The performance of the online and offline greedy algorithms with $\mathcal{D}_1$ as the uniform distribution and $\mathcal{D}_2$ as an arcsine distribution.}
	\label{fig:exp3}
\end{figure}

\begin{figure}[bhpt]
	\centering
	\includegraphics[alt={A plot of proportion of input selected going down and running time going up for two algorithms as input size increases.}, width=0.45\linewidth]{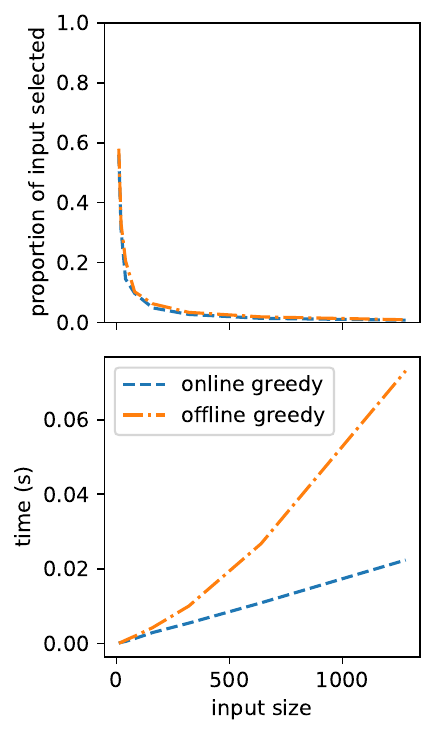}
    \includegraphics[alt={A plot of proportion of input selected going down and running time going down for two algorithms as scale parameter increases.}, width=0.45\linewidth]{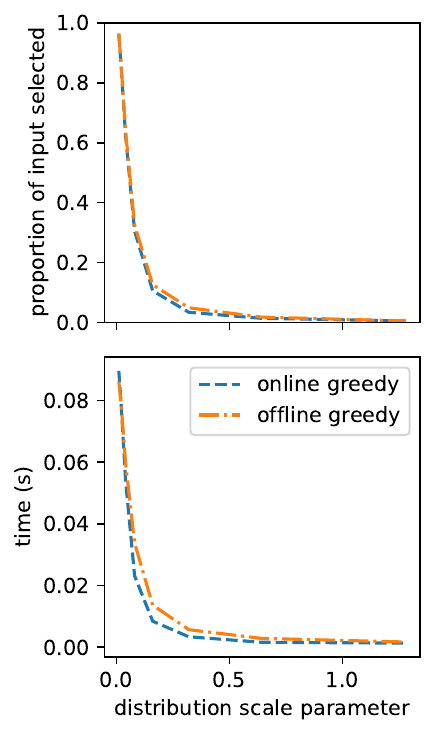}
	\caption{The performance of the online and offline greedy algorithms with $\mathcal{D}_1$ as the uniform distribution and $\mathcal{D}_2$ as a degenerate distribution.}
	\label{fig:exp4}
\end{figure}

For each possible pairing of input distributions, we measure how the proportion of selected rectangles and execution time differs between the two algorithms as either the input size $n$ or scale parameter $s$ varies. The values of $n$ used in our experiments are $\set{10, 20, 40, \dots, 1280}$ with fixed $s = 0.3$. The values of $s$ used are $\set{0.01, 0.02, 0.04, \dots, 1.28}$ with fixed $n = 200$. In all cases, we find that \detgreedy{} is able to nearly match the quality of the solutions found by the offline greedy algorithm while taking significantly less time. The full results of the experiments are shown in Figures~\ref{fig:exp1}--\ref{fig:exp4}.

\clearpage

\section{Additional Example Arrangements}
\label{sec:additional_arrs}
Optimal arrangements of unit hypercubes and unit-volume hyperrectangles for an oblivious adversary against \greedy{p} are shown in Figures~\ref{fig:n3}--\ref{fig:n6}.

\begin{figure}[bhpt]
	\centering
	\begin{tikzpicture}[scale=2]
		\draw (1.5,1.5) rectangle node {1} (2.5,2.5);
		\draw (3.3,0.7) rectangle node {2} (2.3,1.7);
		\draw (0.7,3.3) rectangle node {3} (1.7,2.3);
	\end{tikzpicture}
	\caption{An optimal arrangement of $n=3$ unit squares for $p=0.75$.}
	\label{fig:n3}
\end{figure}
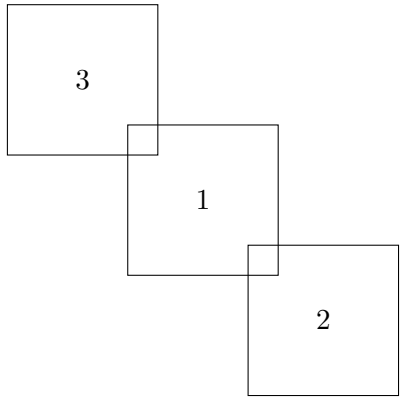

\begin{figure}[bhpt]
	\centering
	\begin{tikzpicture}[scale=2]
		\draw (1.5,1.5) rectangle node {1} (2.5,2.5);
		\draw (3.3,0.7) rectangle node {2} (2.3,1.7);
		\draw (0.7,3.3) rectangle node {3} (1.7,2.3);
		\draw (3.3,3.3) rectangle node {4} (2.3,2.3);
	\end{tikzpicture}
	\caption{An optimal arrangement of $n=4$ unit squares for $p=0.67$.}
	\label{fig:n4}
\end{figure}

\begin{figure}[bhpt]
	\centering
	\begin{tikzpicture}[scale=2]
		\draw (-0.1,-0.1) rectangle node [shift={(-0.2,-0.7)}] {1} (0.9,0.9);
		\draw (0.2,0.2) rectangle node [shift={(-0.2,0.7)}] {2} (1.2,1.2);
		\draw (1.8,-0.7) rectangle node {3} (0.8,0.3);
		\draw (-0.7,1.8) rectangle node {4} (0.3,0.8);
		\draw (1.8,1.8) rectangle node [shift={(0.2,0.2)}] {5} (0.8,0.8);
	\end{tikzpicture}
	\caption{An optimal arrangement of $n=5$ unit squares for $p=0.56$.}
	\label{fig:n5}
\end{figure}

\begin{figure}[bhpt]
\centering
\begin{tikzpicture}[scale=2]
	\draw (-0.1,-0.1) rectangle node [shift={(0.2,-0.7)}] {1} (0.9,0.9);
	\draw (0.2,0.2) rectangle node [shift={(-0.2,0.7)}] {2} (1.2,1.2);
	\draw (1.8,-0.7) rectangle node {3} (0.8,0.3);
	\draw (-0.7,1.8) rectangle node {4} (0.3,0.8);
	\draw (1.8,1.8) rectangle node [shift={(0.2,0.2)}] {5} (0.8,0.8);
	\draw (-0.7,-0.7) rectangle node [shift={(-0.2,-0.2)}] {6} (0.3,0.3);
\end{tikzpicture}
\caption{An optimal arrangement of $n=6$ unit squares for $p=0.50$.}
\label{fig:n6}
\end{figure}

\end{document}